\documentclass[prb,twocolumn,superscriptaddress,amsmath,amssymb,showpacs,floatfix,preprintnumbers]{revtex4}
\usepackage{amsmath}
\usepackage{graphicx}
\usepackage{dcolumn}
\usepackage{color}
\usepackage{lipsum}
\DeclareGraphicsExtensions{.png .jpg .pdf}

\usepackage{dsfont}
\begin{document}

\title{Multilayers black phosphorus: from tight-binding to continuum description}

\author{D. J. P. de Souza}\email{duarte.j@fisica.ufc.br}
\affiliation{Departamento de F\'isica, Universidade Federal do Cear\'a, Caixa Postal 6030, Campus do Pici, 60455-900 Fortaleza, Cear\'a, Brazil}
\author{L. V. de Castro}\email{vieiradecastroluan@fisica.ufc.br}
\affiliation{Departamento de F\'isica, Universidade Federal do Cear\'a, Caixa Postal 6030, Campus do Pici, 60455-900 Fortaleza, Cear\'a, Brazil}
\author{D. R. da Costa}\email{diego_rabelo@fisica.ufc.br}
\affiliation{Departamento de F\'isica, Universidade Federal do Cear\'a, Caixa Postal 6030, Campus do Pici, 60455-900 Fortaleza, Cear\'a, Brazil}
\affiliation{Department of Electrical \& Computer Engineering, University of Minnesota, Minneapolis, Minnesota 55455, USA}
\author{J. Milton Pereira Jr.}\email{pereira@fisica.ufc.br}
\affiliation{Departamento de F\'isica, Universidade Federal do Cear\'a, Caixa Postal 6030, Campus do Pici, 60455-900 Fortaleza, Cear\'a, Brazil}
\author{Tony Low}\email{tlow@umn.edu}
\affiliation{Department of Electrical \& Computer Engineering, University of Minnesota, Minneapolis, Minnesota 55455, USA}

\date{Submitted to Phys. Rev. B: 14 July 2017}

%\date{ \today }

\begin{abstract}
We investigate the electronic properties of $N$-layer black phosphorus by means of an analytical method based on a recently proposed tight-binding Hamiltonian involving $14$ hopping parameters. The method provides simple and accurate general expressions for the Hamiltonian of $N$-layer phosphorene, which are suitable for the study of electronic transport and optical properties of such systems, and the results show the features that emerge as the number of layers increases. In addition, we show that the $N$-layer problem can be translated into $N$ effective monolayer problems in the long wavelength approximation and, within this analytical picture, we obtain expressions for the energy gap and the effective masses for electrons and holes along the $N$-layer black phosphorus plane directions as function of the number of layers, as well as for the Landau levels as function of perpendicular magnetic field.
\end{abstract}

%\textit{Submitted to Phys. Rev. B: 14 July 2017}
\pacs{71.10.Pm, 73.22.-f, 73.63.-b}
\maketitle

\section{Introduction}

The search for new materials with useful electronic properties has led to an increasing interest on the investigation of a class of layered solids that can be produced as single or few layers. These new two-dimensional (2D) materials, which were first brought to attention by the production of graphene in $2004$,\cite{CastroNetoReview, Misha1, Geim} have been shown to display properties that are not found in their bulk form.\cite{silicene, germanene, MoS2, TonyBook, Two-dimensionalAtomicCrystals} Among these substances, there has been considerable interest on the study of black phosphours (BP), an allotrope of phosphorus. \cite{bp1, bp2, bp3, bp4, bp5, bp6, bp7} In contrast with graphene, BP is a semiconductor, and its high electronic mobility makes it a possible candidate for device applications. \cite{bp1, bp2, Yuan, Fazzio} One important aspect of the electronic structure of BP is the dependence of the gap on the number of layers. Experiments have found a band gap in the range of $1.8$ eV for single layer BP which is reduced to $\approx 0.4$ eV for bulk samples. \cite{bp2, bp5, Tran, Gomez, Dolui, Das, Kim, Katsnelson, Andrey}

Recently, a series of calculations have obtained the band structure of BP, both from a first principles approaches,\cite{Tran, Dolui, Rudenko, Tran1, Carvalho, DFTgap} ${\mathbf k} \cdot {\mathbf p}$ methods,\cite{bp6, Zhou, Li, Roberto, Tony} as well as tight-binding \cite{Rudenko, Rudenko1} and continuum models\cite{Milton, DuarteLuan}. The results have shown that BP presents a large anisotropic effective mass and, in addition, that the gap itself can be modified by the application of an external bias.\cite{Katsnelson, Milton} Most of these works have considered single or bilayer BP, due to the increasing computational demands as the number of layers is increased. In this work, we extend the previous proposed tight-binding\cite{Rudenko, Rudenko1} and continuum\cite{Milton} approaches to consider BP films with arbitrary numbers of layers. We show that a system of N coupled BP layers can be approximately mapped into a system of N uncoupled single layers. Expressions for the low-energy electron and hole bands, as well as their effective masses are derived. This in turn permits a straightforward calculation of the Landau level spectrum of the system, as will be also discussed here. 

The paper is organized as follows. In Sec.~\ref{sec.TBmodel}, we present the model Hamiltonian used to describe the charge carriers in single layer (\ref{subsec.TBmono}) and bilayer BP (\ref{subsec.TBbi}), as well as the analytical expressions for the electronic band strcutures, energy gap and effective masses. In Sec. \ref{subsec.TBnlayer}, we generalize the discussion to the case of N-layer BP systems. The continuum approximation is developed in Sec.~\ref{sec.ContinuumModel} and a brief investigation about the Landau level spectrum is explored in Sec.~\ref{sec.LLs}. Finally, in Sec.~\ref{sec.Conclusions} we report the concluding remarks. 

\section{Tight-binding model}\label{sec.TBmodel}

\subsection{Monolayer phosphorene}\label{subsec.TBmono}

\begin{figure}[t]
\centerline{\includegraphics[width = \linewidth]{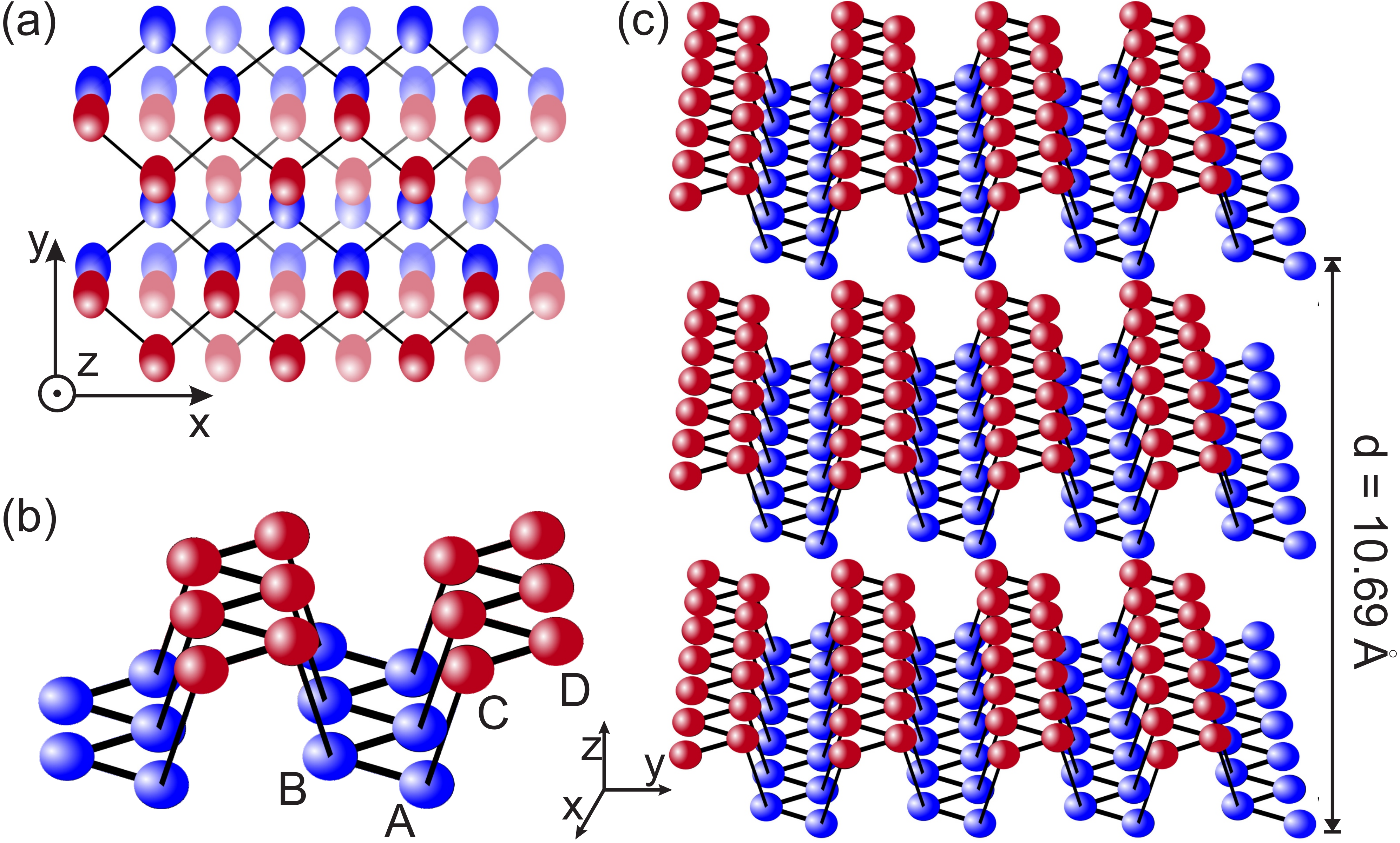}}
\caption{(Color online) Schematic lattice structure of $N$-layer BP system, where the phosphorus atoms at different sublayers are represented by different colors in each monolayer BP. (a) Top view of a N-layer system, emphasizing the AB-stacking and the orientations of the lattice adopted in this work. (b) Side view of a monolayer BP, indicating the four sublattices: $A$ and $B$ at bottom sublayer, and $C$ and $D$ at top sublayer. (c) Side view of a $N$-layer BP system, where $d = 10.69$ \AA\ is the lattice parameter in the out-of-plane direction ($z$-direction).} 
\label{Fig1}
\end{figure}

Figure~\ref{Fig1} illustrates the orthorhombic crystal structure of a N-layer BP system, emphasizing the in-plane orientation adopted in this work and the assumed stacking of the layers (Fig.~\ref{Fig1}(a)). The four inequivalent sublattices in the unit cell and the lattice parameter along the out-of-plane direction are sketched in Fig.~\ref{Fig1}(b) and \ref{Fig1}(c), respectively. The phosphorus atoms at different sublayers are represented by different colors: sublattices $A$ and $B$ ($C$ and $D$) at the bottom (top) sublayer are represented by blue (red) symbols.

The Hamiltonian proposed in Ref.~[\onlinecite{Rudenko1}] for monolayer phosphorene, within the ten-hopping parameter tight-binding approach, in momentum space is given by 
\begin{equation}
H_{mono} = \left(
\begin{array}{cc}
H_0 & H_2 \\
H_2 & H_0
\end{array}\right),
\label{eq1}
\end{equation}
with the following definitions
\begin{subequations}
\begin{align}
H_{0} &= \left(
\begin{array}{cc}
t_{AA}(k) & t_{AB}(k) \\
t_{AB}^{*}(k) & t_{AA}(k)
\end{array}\right),\label{eq2}\\
H_{2} &= \left(
\begin{array}{cc}
t_{AD}(k) & t_{AC}(k) \\
t_{AC}^{*}(k) & t_{AD}(k)
\end{array}\right),\label{eq3}
\end{align}
\end{subequations}
where the structure factors are given in the Appendix \ref{appA}, as well as a schematic view of the intralayer hopping parameters, lattice distances and bond angles is depicted in Figs.~\ref{FigAppendix}(a) and \ref{FigAppendix}(b). The eigenstates of Hamiltonian (\ref{eq1}) are four-component spinors $\Phi = [\phi_A \quad \phi_B \quad \phi_D \quad \phi_C]^{T}$, where the functions $\phi_{A, B, C, D}$ are the probability amplitudes for finding electrons on the atomic sites $A$, $B$, $C$ and $D$. Following Ref.~[\onlinecite{Milton}], we can perform an unitary transformation to rewrite the monolayer Hamiltonian in a simpler block form. The new Hamiltonian and eigenstates are given by
\begin{equation}
H_k^{'} = \left(
\begin{array}{cc}
H_{k}^{+} & 0 \\
0 & H_{k}^{-}
\end{array}\right), \quad\mbox{and}\quad
\psi_{k}' = \left(
\begin{array}{c}
\Psi_{k}^{+}  \\
\Psi_{k}^{-} 
\end{array}\right),
\label{eq4}
\end{equation}
where
\begin{equation}
H_k^{\pm} = H_{0} \pm H_{2}, \quad\mbox{and}\quad
\Psi_{k}^{\pm} = \left(
\begin{array}{c}
\phi_A \pm \phi_D  \\
\phi_B \pm \phi_C 
\end{array}\right).
\label{eq5}
\end{equation}
By diagonalizing the Hamiltonian~(\ref{eq5}), one obtains the following energy bands
\begin{equation}
E_{s}^{\pm}(k) = t_{AA}(k) \pm t_{AD}(k) + s |t_{AB}(k) \pm t_{AC}(k)|,
\label{eq6}
\end{equation}
where $s = \pm$ denotes the valence ($+$) and conduction ($-$) bands. A more detailed analysis of Eq.~(\ref{eq6}) reveals that the bands associated to $E_{s}^{+}(k)$ have lower energies than the bands associated with $E_{s}^{-}(k)$. In Fig.~\ref{bands}(a), we plot the low energy bands ($E_{s}^{+}(k)$) given by Eq.~(\ref{eq6}) with solid blue curve. Therefore, the energy gap for the monolayer BP is obtained by the eigenstates of $H_{k}^{+}$ sub-Hamiltonian, namely $E_{g}^{mono} = 2|t_{AB}(0) + t_{AC}(0)| \approx 1.838$ eV.

\subsection{Bilayer phosphorene}\label{subsec.TBbi}

For bilayer BP, we have to incorporate the coupling between adjacent layers, which are separated by a distance of $\approx 3.214$ \AA\ and consider eight sublattices. A sketch of the four interlayer hopping parameters is shown in Fig.~\ref{FigAppendix}(c) in Appendix. The additional coupling terms were already computed in Ref.~[\onlinecite{Milton}] in the context of the five-intralayer-hopping parameters tight-binding approach. The generalization to the ten-intralayer-hopping parameters tight-binding model is straightforward, since both approximations have the same definitions for the interlayer hoppings, differentiating only by the values of the hopping parameters.\cite{Rudenko, Rudenko1} Therefore, according to Ref.~[\onlinecite{Milton}], we can write the Hamiltonian and eigenstates for bilayer BP as
\begin{equation}
H_{bi} = \left(
\begin{array}{cc}
H & H_{c} \\
H_{c}^{\dagger} & H
\end{array}\right), \quad\mbox{and}\quad \Psi_{bi} =
\left(
\begin{array}{c}
\Phi_{1}  \\
\Phi_{2}
\end{array}\right),
\label{eq7}
\end{equation}
with  $\Phi_{i} = [\phi_{A,i} \qquad \phi_{B,i} \qquad \phi_{D,i} \qquad \phi_{C,i}]^{T}$, where $i = 1, 2$ is the layer index. $H$ is the Hamiltonian associated with each monolayer BP and $H_{c}$ contains the contribution of the couplings between atomic sites located in adjacent layers, which is given by
\begin{equation}
H_{c} \hspace{-0.025cm}=\hspace{-0.025cm} \left(\hspace{-0.05cm}
\begin{array}{cc}
0 & H_{3} \\
0 & 0
\end{array}\hspace{-0.05cm}\right),\hspace{0.05cm}\mbox{with}\hspace{0.2cm}
H_{3} \hspace{-0.025cm}= \hspace{-0.025cm}\left(\hspace{-0.05cm}
\begin{array}{cc}
t_{AD'}(k) & t_{AC'}(k) \\
t_{AC'}^{*}(k) & t_{AD'}(k)
\end{array}\hspace{-0.05cm}\right).
\label{eq8}
\end{equation}
The interlayer structure factors in $H_3$ are defined in the Appendix \ref{appA}.

Similar to the case of the monolayer BP, we can perform an unitary transformation to rewrite the bilayer Hamiltonian (\ref{eq7}) in simpler form, in order to avoid to deal with eight coupled equations. Assuming a bias perpendicular to the bilayer plane in such a way that the on-site energy for the atoms at the top (bottom) monolayer is $\Delta /2$ ($-\Delta /2$) and applying the unitary transformation presented in Appendix \ref{appB}, we arrive at the following Hamiltonian for bilayer BP as
\begin{equation}
H_{k}^{\pm} = \left(
\begin{array}{cc}
H_0 \pm H_2 +  H_3 / 2 &  \Delta/2 \\
\Delta/2  & H_0 \pm H_2 - H_3 / 2
\end{array}\right),
\label{eq9}
\end{equation}
where the Hamiltonian $H_{k}^{+}$ ($H_{k}^{-}$) describes the low (high) energy bands.
The eigenstates of $H_ {k}^{\pm}$ are
\begin{equation}
\Psi_{k}^{\pm} = \frac{e^{i\theta_\pm}}{2}\left(
\begin{array}{c}
(\phi_{A,1} \pm \phi_{D,1}) + (\phi_{A,2} \pm \phi_{D,2})  \\
(\phi_{B,1} \pm \phi_{C,1}) + (\phi_{B,2} \pm \phi_{C,2})  \\
(\phi_{A,1} \pm \phi_{D,1}) - (\phi_{A,2} \pm \phi_{D,2})  \\
(\phi_{B,1} \pm \phi_{C,1}) - (\phi_{B,2} \pm \phi_{C,2})  \\
\end{array}\right),
\label{eq10}
\end{equation}
with $\theta_+ = 0$ and $\theta_- = -\pi/2$. Therefore, diagonalizing Hamiltonian (\ref{eq9}), we obtain the following expressions for the energy bands close to the Fermi level 
\begin{subequations}
\begin{align}
E_c \hspace{-0.075cm} &= \hspace{-0.075cm}\frac{\epsilon_1^{+}\hspace{-0.075cm} +\hspace{-0.075cm} \epsilon_2^{+}\hspace{-0.075cm} +\hspace{-0.075cm} \epsilon_1^{-}\hspace{-0.075cm} +\hspace{-0.075cm} \epsilon_2^{-}}{2}\hspace{-0.075cm} \pm \hspace{-0.075cm}\sqrt{\hspace{-0.05cm}\left[\hspace{-0.05cm}\frac{\epsilon_1^{+}\hspace{-0.075cm} +\hspace{-0.075cm} \epsilon_2^{+} \hspace{-0.075cm}-\hspace{-0.075cm} \epsilon_1^{-}\hspace{-0.075cm} -\hspace{-0.075cm} \epsilon_2^{-}}{2}\hspace{-0.05cm}\right]^{2}\hspace{-0.075cm} +\hspace{-0.075cm} \left(\hspace{-0.05cm}\frac{\Delta}{2}\hspace{-0.05cm}\right)^{2}},\label{eq11}\\
E_v \hspace{-0.075cm}&= \hspace{-0.075cm}\frac{\epsilon_1^{+}\hspace{-0.075cm} - \hspace{-0.075cm}\epsilon_2^{+}\hspace{-0.075cm} + \hspace{-0.075cm}\epsilon_1^{-} \hspace{-0.075cm} -\hspace{-0.075cm} \epsilon_2^{-}}{2}\hspace{-0.075cm} \pm \hspace{-0.075cm}\sqrt{\hspace{-0.05cm} \left[\hspace{-0.05cm}\frac{\epsilon_1^{+}\hspace{-0.075cm} -\hspace{-0.075cm} \epsilon_2^{+} \hspace{-0.075cm}-\hspace{-0.075cm} \epsilon_1^{-}\hspace{-0.075cm} +\hspace{-0.075cm} \epsilon_2^{-}}{2}\hspace{-0.05cm}\right]^{2}\hspace{-0.075cm} +\hspace{-0.075cm} \left(\hspace{-0.05cm}\frac{\Delta}{2}\hspace{-0.05cm}\right)^{2}},\label{eq12}
\end{align}
\end{subequations}
corresponding to the conduction ($E_c$) and valence ($E_v$) bands, where the functions of the wavevector $\epsilon_{i}^{\pm}$, with $i = 1, 2$, are defined by 
\begin{subequations}
\begin{align}
\epsilon_{1}^{\pm} &= t_{AA}(k) + t_{AD}(k) \pm t_{AD'}(k)/2, \\
\epsilon_{2}^{\pm} &= |t_{AB}(k) + t_{AC}(k) \pm t_{AC'}(k)/2|. 
\label{eq13}
\end{align}
\end{subequations}
The plot of the energy bands given by Eqs.~(\ref{eq11}) and (\ref{eq12}) is depicted by solid blue curve in Fig.~\ref{bands}(b). The energy gap of the bilayer BP is given by $E_{g}^{bi} = 2|t_{AB}(0) + t_{AC}(0) - t_{AC'}(0)/2| \approx 1.126$ eV. Note that for the zero bias case ($\Delta = 0$), the bilayer Hamiltonian (\ref{eq9}) has the same form as the monolayer Hamiltonian (\ref{eq4}), except for the modified diagonal matrix elements due to the presence of interlayer coupling term $H_3$. 

\subsection{N-layer phosphorene}\label{subsec.TBnlayer}

In this section, we generalize the previous discussions to the case of $N$-layer BP. The layers are stacked according to the configuration sketched in Fig.~\ref{Fig1}(a), which is called AB stacking.\cite{Peeters, Rahaman} The Hamiltonian for a $N$-layer system follows from a natural generalization of Hamiltonian (\ref{eq7}), given by
\begin{equation}
H_{N} = \left(\hspace{-0.05cm}
\begin{array}{ccccc}
H & H_c &  &  & \\
H_c^{\dagger} & H & H_c & \\
  & H_c^{\dagger} & H & H_c  & \\
 &  &  & \ddots & \\
  &  & & & H_{c}\\
& & \qquad & H_c^{\dagger} & H 
\end{array}\hspace{-0.05cm}\right)_{N\times N},
\label{eq14}
\end{equation}
which acts on the eigenstate
\begin{equation}
\Psi_{N} = \left(
\begin{array}{c}
\Phi_{1} \\
\Phi_{2}\\
 \vdots\\
\Phi_{N}\\ 
\end{array}\right)_{N\times 1}, \quad\mbox{with}\quad
\Phi_{i} = \left(
\begin{array}{c}
\phi_{A,i} \\
\phi_{B,i}\\
\phi_{D,i}\\
\phi_{C,i}\\ 
\end{array}\right).
\label{eq15}
\end{equation}
One can notice that the Hamiltonian (\ref{eq14}) is a tridiagonal matrix formed by $4\times 4$ blocks, since we only consider the coupling between the adjacent layers, otherwise off-tridiagonal terms would be non-null. The main diagonal of $H_{N}$ is composed by monolayer type Hamiltonians, similar to Eq.~(\ref{eq1}), and the adjacent diagonals are populated by $H_c$ blocks that contain the interaction terms connecting sublattice sites between the adjacent layers, similar to the coupling matrix of the bilayer case given by Eq.~(\ref{eq8}). This corresponds to the case of a $N$-layer BP system that is free of interactions with any external sources, as for instance external electric and magnetic fields, which can be easily incorporated in the following formalism through perturbation theory.

The eigenvalue equation $H_{N} \Psi_{N} = E\Psi_{N}$ is equivalent to a set of equations of the form 
\begin{equation}
H_{c}^{\dagger}\Phi_{i - 1} + (H - E)\Phi_{i} + H_{c}\Phi_{i + 1} = 0,
\label{eq16}
\end{equation}
obeying the boundary condition $\Phi_{0} = \Phi_{N+1} = 0$. $i = 1, 2, ... , N$ is the layer index. Eq.~(\ref{eq16}) can be equivalently rewritten in the following pair of equations for each $i$ 
\begin{subequations}
\begin{align}
&(H_0 - E)\psi_{AB,i} + H_2 \psi_{DC,i} + H_3 \psi_{DC,i + 1} = 0, \label{eq17a}\\
&(H_0 - E)\psi_{DC,i} + H_2 \psi_{AB,i} + H_3 \psi_{AB,i - 1} = 0, \label{eq17b}
\end{align}
\end{subequations}
where the two sets of two-component spinors are defined by $\psi_{AB,i} = [\phi_{A,i} \quad \phi_{B,i}]^{T}$ and $\psi_{DC,i} = [\phi_{D,i} \quad \phi_{C,i}]^{T}$. Therefore, we have separated the amplitudes for each sublayer of each monolayer $i$, i.e. $\psi_{AB,i}$ ($\psi_{DC,i}$) contains separately the amplitudes for the bottom (top) sublayer of the $i$-th monolayer. Before proceeding, it is necessary to comment on very important points related to the sublattice amplitudes and the energy levels of the $N$-layer BP system. As a means to it, we shall exemplify these features through the monolayer and bilayer BP cases. 

As discussed previously in Sec.~\ref{subsec.TBmono}, the $H_{k}^{+}$ ($H_{k}^{-}$) Hamiltonian is associated with low (high) energy bands around the Fermi level. We can interpret this feature by analyzing the eigenstates associated with each sub-Hamiltonian. The eigenstate $\Psi_{k}^{+}$ ($\Psi_{k}^{-}$) of the sub-Hamiltonian $H_{k}^{+}$ ($H_{k}^{-}$) is given by the sum (difference) of the probability amplitudes of the equivalent sublattices for each component, i.e. $\Psi_{k}^{\pm} = \psi_{AB} \pm \psi_{DC}$. In other words, the lowest (highest) energy bands are associated with (anti-)bonds states between each sublayer. meaning $\psi_{AB} + \psi_{DC}$ ($\psi_{AB} - \psi_{DC}$). This is analogous to the case of the hydrogen molecule, where the bond state has lower energy than the anti-bond state. Thus, we use the nomenclature bond and anti-bond in the context of phosphorene as an analogy with the molecular orbital theory. 

\begin{figure}[t]
\centerline{\includegraphics[width = \linewidth]{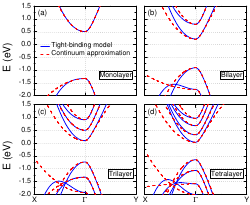}}
\caption{(Color online) Band structures for (a) monolayer, (b) bilayer, (c) trilayer and (d) tetralayer phosphorene obtained by using the analytical expressions Eqs.~(\ref{eq23}) and (\ref{eq44}) within tight-binding model (blue solid curves) and continuum approximation (red dashed curves), respectively.} 
\label{bands}
\end{figure}

Similar feature is also observed for the bilayer BP case (see Eqs.~(\ref{eq9}) and (\ref{eq10})). The Hamiltonian that describes low (high) energy excitations, $H_{k}^{+}$ ($H_{k}^{-}$), has eigenstates that are bond (anti-bond) states between each sublayer in the adjacent layer, as can be seen in Eq.~(\ref{eq10}). However, for the bilayer BP case, one has an additional feature: their eigenstates have not only bonds and anti-bonds between the sublayers in each layer, but also they have bonds and anti-bonds between the adjacent layers. Here, the bond states between different layers [$(\psi_{AB,1} + \psi_{DC,1}) + (\psi_{AB,2} + \psi_{DC,2})$] exhibit lower energies than the anti-bond ones [$(\psi_{AB,1} + \psi_{DC,1}) - (\psi_{AB,2} + \psi_{DC,2})$]. Therefore, regarding this analysis we can predict four low energy bands for bilayer BP, which is in accordance with the previous section and Fig.~\ref{bands}(b).

This argument can be generalized to the multilayer BP case. The lower energy bands around the energy gap are described by bonds of $\psi_{AB,i}$ and $\psi_{DC,i}$ for each layer $i$, i.e. $\psi_{AB,i} + \psi_{DC,i}$, whereas the higher energy band located away from the gap region are described by anti-bonds of $\psi_{AB,i}$ and $\psi_{DC,i}$ for each layer $i$, i.e. $\psi_{AB,i} - \psi_{DC,i}$. This relation between the energy bands and the bond states allows us to separate the high and low energy excitations in a very intuitive way. Rewriting Eqs.~(\ref{eq17a}) and (\ref{eq17b}) in the basis of bonding and anti-bonding amplitudes, we arrive at
\begin{align}
&(H_0 \hspace{-0.05cm}\pm\hspace{-0.05cm} H_2 \hspace{-0.05cm}-\hspace{-0.05cm} E)(\psi_{AB,1} \hspace{-0.05cm}\pm\hspace{-0.05cm} \psi_{DC,1}) \hspace{-0.05cm}+\hspace{-0.05cm} H_3 \psi_{DC,2} \hspace{-0.05cm}=\hspace{-0.05cm} 0, \nonumber \\
&(H_0 \hspace{-0.05cm}\pm\hspace{-0.05cm} H_2 \hspace{-0.05cm}-\hspace{-0.05cm} E)(\psi_{AB,2} \hspace{-0.05cm}\pm\hspace{-0.05cm} \psi_{DC,2}) \hspace{-0.05cm}+\hspace{-0.05cm} H_3 (\psi_{DC,3} \hspace{-0.05cm}\pm\hspace{-0.05cm} \psi_{AB,1}) \hspace{-0.05cm}=\hspace{-0.05cm} 0, \nonumber\\
&\hspace{1cm}   \vdots \hspace{2.25cm} \vdots \hspace{3cm}  \vdots \label{eq18}\\
&(H_0 \hspace{-0.05cm}\pm\hspace{-0.05cm} H_2 \hspace{-0.05cm}-\hspace{-0.05cm} E)(\psi_{AB,N} \hspace{-0.05cm}\pm\hspace{-0.05cm} \psi_{DC,N}) \hspace{-0.05cm}\pm\hspace{-0.05cm} H_3 \psi_{AB,N-1} \hspace{-0.05cm}=\hspace{-0.05cm} 0.\nonumber
\end{align}
where the sign $+$ ($-$) denotes the bonding (anti-bonding) states. At this point, we can define the bonding and anti-bonding  orbitals for each layer as $\phi_{i} = \psi_{AB,i} + \psi_{DC,i}$ and $\bar{\phi}_{i} = \psi_{AB,i} - \psi_{DC,i}$, respectively.

In order to take into account just the multilayer BP properties at low energies, we shall reduce the problem in half, i.e. instead of diagonalizing a $N \times N$ block Hamiltonian, we treat only an effective $N/2 \times N/2$ block Hamiltonian, since the low energy bands correspond to the bond states, one can consider only the half of Eqs.~(\ref{eq18}) with the $+$ sign.  

Due to the sublattice symmetry between atomic sites $A/D$ and $B/C$ in each monolayer BP, as a consequence of the $D_{2h}$ group invariance of the BP lattice,\cite{Milton, Ezawa} we regard the following approximation with respect to the sublattice amplitudes: $\psi_{AB,i} \approx \psi_{DC,i}$ for $i = 1, 2, ... , N$, which corresponds to $\phi_{A,i} \approx \phi_{D,i}$ and $\phi_{B,i} \approx \phi_{C,i}$. This is valid for the vast majority of cases of physical interest, since it is very difficult experimentally to induce bias just in one single-layer BP in set of $N$-layer and consequently breaking this sublattice symmetry. On the other hand, it is important to mention that this assumption does not exclude the possibility of applying a perpendicular electric field to the system, assuming that the field affects equally the on-site energy of all atoms in a same layer $i$ by $\epsilon_{i}$. With this in mind, we arrive at the important relation
\begin{equation}
\psi_{AB,i} \approx \psi_{DC,i} \approx \frac{1}{2}\phi_{i}.
\label{eq19}
\end{equation}
Thus, we can rewrite Eq.~(\ref{eq18}) within this approximation in terms of the bonding orbitals, resulting in the following set of equations for $i = 1, 2, ... , N$ BP layers
\begin{equation}
(H_0 + H_2 - E)\phi_{i} + \frac{1}{2}H_3 (\phi_{i - 1} + \phi_{i + 1}) = 0,
\label{eq20}
\end{equation}
obeying the boundary condition $\phi_{0} = \phi_{N+1} = 0$. This boundary condition is satisfied by the following \textit{ansatz}: $\phi_{j} = A\sin(j n\pi/(N + 1))$, where $A$ is a two-component spinor and depends only on $k_{x}$ and $k_{y}$. By taking this \textit{ansatz}, one can easily check that the following identity holds true: $\phi_{i-1} + \phi_{i + 1} = 2\cos(n\pi/(N + 1))\phi_{i}$, where we have used the trigonometrical identity $\sin(a \pm b) = \sin(a)\cos(b) \pm \sin(b)\cos(a)$. A more rigorous way to obtain this identity can be found by the theory of the Toeplitz matrix. It is known that $\sin(jn\pi/(N +1))$ is the $j$-th component of the eigenvector $u$ of the matrix $T$, defined by 
\begin{equation}
T = \left(
\begin{array}{ccccc}
0 & 1 &  & & \\
1 & 0 &1 & \\
 & 1 & 0 & 1  & \\
 &  &  & \ddots & \\
 & &  &  &1 \\
& & \qquad & 1 &0 
\end{array}\right),
\label{eq21}
\end{equation}
with eigenvalues $\lambda_{n} = 2\cos(n\pi/(N+1))$. Thus, the eigenvalue equation $Tu = \lambda u$ results in $u_{j-1} + u_{j+1} = \lambda_n u_{j}$, which is similar to the previous derived identity. 

By substituting the \textit{ansatz} into Eq.~(\ref{eq20}), we obtain
\begin{equation}
\left[H_0 + H_2 + \cos\left(\frac{n\pi}{N+1}\right)H_3\right]\phi_{i}  = E\phi_i,
\label{eq22}
\end{equation}
where $\phi_{i} = \phi_{i}^{n}$ and $n = 1, 2, ..., N$. In summary, the assumed \textit{ansatz} diagonalizes the full Hamiltonian for the systems of $N$ coupled BP layers with the sublattice symmetry approximation within the tight-binding picture. Therefore, we have transformed the complicated problem of diagonalizing a $N\times N$ tridiagonal hermitian block matrix, composed by $4\times 4$ blocks, to $N$ problems of order $2$ for the low energy excitation case. There are $N$ more $2\times 2$ blocks corresponding to the highest energy excitations. The eigenvalues of the Hamiltonian (\ref{eq22}) are easily found to be
\begin{align}
E_n^{\pm}(k, N) \hspace{-0.05cm} &= \hspace{-0.05cm}t_{AA}(k)\hspace{-0.05cm} + \hspace{-0.05cm}t_{AD}(k) \hspace{-0.05cm}+ \hspace{-0.05cm}\cos\left(\hspace{-0.05cm}\frac{n\pi}{N+1}\hspace{-0.05cm}\right)t_{AD'}(k)  \nonumber \\
&\hspace{-1cm}\pm \left|t_{AB}(k)\hspace{-0.05cm} + \hspace{-0.05cm}t_{AC}(k)\hspace{-0.05cm} +\hspace{-0.05cm} \cos\left(\hspace{-0.05cm}\frac{n\pi}{N+1}\hspace{-0.05cm}\right)t_{AC'}(k)\right|.
\label{eq23}
\end{align}
In Fig.~\ref{bands} we show the band structure (solid blue curves) for (a) monolayer, (b) bilayer, (c) trilayer and (d) tetralayer phosphorene obtained from Eq.~(\ref{eq23}) with $N = 1$, $2$, $3$ and $4$, respectively. The agreement with the results found in the literature\cite{Rudenko} is remarkable, showing that the simple sublattice symmetry approximation assumed here (Eq.~(\ref{eq19})) is an excellent and appropriate approach to describe $N$-layer BP systems. 

\begin{figure}[t]
\centerline{\includegraphics[width = 0.9\linewidth]{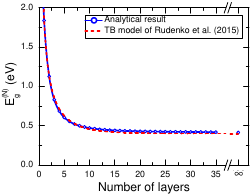}}
\caption{(Color online) Energy gap as a function of the number of layers. The open symbols highlights the behavior of the analytical expression (\ref{eq25}) for integer values of the number of layers $N$. Red dashed curve shows the result obtained by Rudenko et al.\cite{Rudenko1} in order to make a direct comparison with our analytical result.} 
\label{Fig3}
\end{figure} 

The energy gap with an explicit layer-dependence for $N$-layer BP system can also be easily computed from Eq.~(\ref{eq23}), resulting
\begin{equation}
E_g^N = 2 \left| t_{AB}(0) + t_{AC}(0)\hspace{-0.05cm} + \cos\left(\hspace{-0.05cm}\frac{N\pi}{N+1}\hspace{-0.05cm}\right)t_{AC'}(0)\right|,
\label{eq24}
\end{equation}
where the states with $n = N$ are the ones with lowest energies. By applying the identity $\cos(N\pi/(N+1)) = -\cos(\pi/(N+1))$ and recognizing that $E_g^{mono} = 2|t_{AB}(0) + t_{AC}(0)|$ is the energy gap for the monolayer case, we can rewrite the above Eq.~(\ref{eq24}) as
\begin{equation}
E_g^N = \left|E_{g}^{mono}- 2\cos\left(\hspace{-0.05cm}\frac{\pi}{N+1}\hspace{-0.05cm}\right)t_{AC'}(0)\right|.
\label{eq25}
\end{equation}

Figure~\ref{Fig3} displays the energy gap as a function of the number of layers obtained via Eq.~(\ref{eq25}). A comparison with the recently obtained results by Rudenko et al.\cite{Rudenko1} based on the computational analysis of the tight-binding model (red dashed curve in Fig.~\ref{Fig3}) shows an excellent agreement with our analytical result and confirms that the sublattice symmetry approximation assumed here is very accurate and possibly exact within the tight-binding approach. It is worth to mention that, to the best of our knowledge, the energy gap dependence with the number of layers has never been analytically deduced as a natural consequence of a model that describes the electronic properties of $N$-layer BP system. Several approaches have been adopted to circumvent the very difficult calculations imposed by the ten-hopping tight-binding model in order to obtain one single expression for the gap-layer dependence, which is very important for describing optical transition.\cite{Andrey} Most of previous works\cite{Tran, Dolui, Rudenko, Tran1, Carvalho, DFTgap, Rudenko1, PNAS} just numerically fit the data points obtained by first-principles calculations. The usual analytical approach to deal with this problem is based on an approximated quasi-$1$D tight-binding model along the $z$-direction, where each phosphorene layer is associated with an atomic site of $z$-Hamiltonian and the interlayer coupling energy is linked to the hopping parameter of the $1$D chain.\cite{Roberto, Tony, Andrey, DFTgap1} The lack of information about the hopping parameter, i.e. ``interlayer coupling'', is overcome by fitting the \textit{ab-initio} results or experimental data. Here, Eq.~(\ref{eq25}) was obtained from rigorous diagonalization of the Hamiltonian (\ref{eq14}). In the limit of a large number of BP layers ($N \rightarrow \infty$), i.e. within the bulk limit, the energy gap (\ref{eq25}) tends to $0.414$ eV (see last point in Fig.~\ref{Fig3}), which is consistent with the values obtained by previous first-principles calculations\cite{DFTgap} ($\approx 0.43$ eV) and tight-binding model\cite{Rudenko, Rudenko1} ($\approx 0.40$ eV).

Let us now analyze some particular cases ($N=2$ and $N=3$) to explicitly illustrate our general multilayer approximation. For the bilayer case ($N = 2$), the sub-Hamiltonians (\ref{eq22}) reads
\begin{subequations}
\begin{align}
& H_0 + H_2 + \cos\left(\frac{\pi}{3}\right)H_3, \\ 
& H_0 + H_2 + \cos\left(\frac{2\pi}{3}\right)H_3.
\label{eq26}
\end{align}
\end{subequations}
Since $\cos(\pi/3) = - \cos(2\pi/3) = 1/2 $, we arrive at
\begin{equation}
H_{bi} = \left(\hspace{-0.05cm}
\begin{array}{cc}
H_0 + H_2 + H_3 / 2 & 0 \\
0  & H_0 + H_2 - H_3 / 2
\end{array}\hspace{-0.05cm}\right),
\label{eq27}
\end{equation}
which is exactly the same Hamiltonian (Eq.~(\ref{eq9})) found previously for low energy excitations and zero bias ($\Delta = 0$). Keeping in mind that Eq.~(\ref{eq9}) is exact within the tight-binding model, it leads us to an additional indication that the sublattice symmetry approximation is exact for zero bias bilayer BP. For the trilayer BP ($N = 3$), we have
\begin{subequations}
\begin{align}
& H_0 + H_2 + \cos\left(\frac{\pi}{4}\right)H_3, \\ 
& H_0 + H_2 + \cos\left(\frac{2\pi}{4}\right)H_3, \\
& H_0 + H_2 + \cos\left(\frac{3\pi}{4}\right)H_3.
\label{eq28}
\end{align}
\end{subequations}
Since $\cos(\pi/4) = - \cos(2\pi/4) = 1/\sqrt{2} $ and $\cos(2\pi/4) = 0$, the effective Hamiltonian for low-energy excitations in trilayer BP becomes
\begin{equation}
H_{tri} \hspace{-0.1cm}=\hspace{-0.1cm} \left(\hspace{-0.1cm}
\begin{array}{ccc}
H_0\hspace{-0.1cm} + \hspace{-0.1cm}H_2 \hspace{-0.1cm}+\hspace{-0.1cm} H_3 / \sqrt{2} & 0 & 0 \\
0  & H_0 + H_2  & 0 \\
0 & 0 &H_0 \hspace{-0.1cm}+\hspace{-0.1cm} H_2 \hspace{-0.1cm}-\hspace{-0.1cm} H_3 / \sqrt{2} 
\end{array}\hspace{-0.1cm}\right).
\label{eq29}
\end{equation}
It is interesting to note that the effective Hamiltonian for the trilayer BP (\ref{eq29}) is composed by one monolayer Hamiltonian and one ``bilayer Hamiltonian'', except for a factor of $1/\sqrt{2}$ instead of $1/2$ in bilayer terms. This implies that trilayer and monolayer phoephorene share energy bands, as can also be seen in Figs.~\ref{bands}(a) and \ref{bands}(c). The same feature is also observed for the multilayer case whenever the number of layers is odd. A sub-Hamiltonian for the $N$-layer case, $H_{n} = H_0 + H_2 + \cos(n\pi/(N+1))H_3$, is a monolayer Hamiltonian only when $\cos(n\pi/(N+1)) = 0$ for some $n$ [see Eq.~(\ref{eq5})]. When the number of layers is odd, i.e. $N = 2m + 1$ with $m = 1,2,...$, we find that $n = m + 1$ generates a monolayer Hamiltonian. For instance in the trilayer BP case, we have $m = 1$, implying that the sub-Hamiltonian for $n = 2$ is a monolayer type. Therefore, we can always find one monolayer type sub-Hamiltonian $H_{n = m + 1}$. On the other hand, if the number of layers is even, i.e. $N = 2m$, the condition for a monolayer type Hamiltonian would be $n = m + 1/2$, but it is never satisfied, since $m$ and $n$ are both integers. In summary, the N-layer Hamiltonian is composed by $N/2$ bilayer Hamiltonians if $N$ is even, and $(N-1)/2$ bilayer and one monolayer Hamiltonians if $N$ is odd. 

Similar features hold true for high energy bands, but now the basis is constituted by anti-bonding orbitals for each monolayer BP, instead of bond states, as previously discussed. Hence, one has to assume a lattice antisymmetry approximation, instead of symmetric one, such as: $\psi_{AB,i} \approx -\psi_{DC,i} \approx \bar{\phi_{i}}/2$, where $\bar{\phi_{i}} = \psi_{AB,i} - \psi_{DC,i}$ are the anti-bonding orbitals. Therefore, we can generalize our results by stating that the $N$-layer BP Hamiltonian for any energy range within the ten-hopping tight-binding model has a diagonal form in which each sub-Hamiltonian is given by
\begin{equation}
H_{n}^{\pm} = H_0 \pm H_2 \pm \cos\left(\frac{n\pi}{N+1}\right)H_3,
\label{eq30} 
\end{equation} 
where the sign $+$ ($-$) corresponds to the low (high) energy bands and $n = 1, 2, ..., N$ is the subband index. It gives a total of $2N$ matrix equations of order $2$, which is equivalent to $N$ equations of order $4$ or one matrix equation of order $4N$ similar to the initial Hamiltonian (\ref{eq14}).

\section{Continuum Approximation}\label{sec.ContinuumModel}

Despite the significant simplification for $N$-layer BP Hamiltonian (\ref{eq14}), given by Eq.~(\ref{eq30}), the structure factors for the ten-hopping parameters tight-binding model are still not tractable for analytical investigation of the electronic properties away from the $\Gamma$ point. Thus, further simplification is desirable in order to make the resulting model more suitable for analytical calculations. Within the long-wavelength approximation, a simple analytical model can be derived by expanding the structure factors (see Appendix~\ref{appA}) up to second order in $k$. It has been recently shown within the five-hopping parameter approach\cite{Milton, DuarteLuan} that this continuum approximation is very suited for describing the physics of large BP systems, yielding very accurate results within its limit of validation. Moreover, its applicability is not restricted to monolayer case, but it can be extended to multilayer BP, being this way less computationally demanding than tight-binding model and first-principles calculations.   

By expanding the structure factors, given in Appendix~\ref{appA}, around the $\Gamma$ point up to second order in $k$, one obtains the following expressions
\begin{subequations}
\begin{align}
t_{AA} &= \delta_{AA} + \eta_{AA}k_{x}^{2} + \gamma_{AA}k_y^{2},\label{31}\\
t_{AB} &= \delta_{AB} + \eta_{AB}k_{x}^{2} + \gamma_{AB}k_y^{2} + i\chi_{AB}k_y,\label{32}\\
t_{AC} &= \delta_{AC} + \eta_{AC}k_{x}^{2} + \gamma_{AC}k_y^{2} + i\chi_{AC}k_y,\label{33}\\
t_{AD} &= \delta_{AD} + \eta_{AD}k_{x}^{2} + \gamma_{AD}k_y^{2},\label{34}
\end{align}
\end{subequations}
for the intralayer terms and
\begin{subequations}
\begin{align}
t_{AC'} &= \delta_{AC'} + \eta_{AC'}k_{x}^{2} + \gamma_{AC'}k_y^{2} + i\chi_{AC}k_y,\label{35}\\
t_{AD'} &= \delta_{AD'} + \eta_{AD'}k_{x}^{2} + \gamma_{AD'}k_y^{2},\label{36}
\end{align}
\end{subequations}
for the interlayer contributions. The coefficient values of the expanded structure factors (Eqs.~(\ref{31})-(\ref{36})) for both five-hopping and ten-hopping models are summarized in Table I.
\begin{table}[t]
\label{table1}
\caption{Structure factor coefficients for both five and ten-hopping continuum approximation.}
\centering 
\begin{tabular}{cccc}
\hline\hline
& \hspace{0.25cm}10-hopping\hspace{0.25cm} & \hspace{0.25cm}5-hopping\hspace{0.25cm} & \hspace{0.25cm}units \\
\hline
$\delta_{AA}$ & -0.338 & 0.00 & eV \\
$\delta_{AB}$ & -2.912 & -2.85 & eV\\
$\delta_{AC}$ & 3.831 & 3.61 & eV\\
$\delta_{AD}$ & -0.076 & -0.42 & eV\\
\hline
$\delta_{AC'}$ & 0.712 & 0.41 & eV \\
$\delta_{AD'}$ & -0.132 & -0.06 & eV \\
\hline
$\eta_{AA}$ & 1.161 & 0.00 & eV$\cdot$\AA$^{2}$ \\
$\eta_{AB}$ & 2.05 & 3.91 & eV$\cdot$\AA$^{2}$ \\
$\eta_{AC}$ & 0.460 & -0.53 & eV$\cdot$\AA$^{2}$ \\
$\eta_{AD}$ & 0.104 & 0.58 & eV$\cdot$\AA$^{2}$ \\
\hline
$\eta_{AC'}$ & -0.9765 & -0.56 & eV$\cdot$\AA$^{2}$ \\
$\eta_{AD'}$ & 2.699 & 3.31 & eV$\cdot$\AA$^{2}$ \\
\hline
$\gamma_{AA}$ & -1.563 & 0.00 & eV$\cdot$\AA$^{2}$\\
$\gamma_{AB}$ & 3.607 & 4.41 & eV$\cdot$\AA$^{2}$ \\
$\gamma_{AC}$ & -1.572 & 0.00 & eV$\cdot$\AA$^{2}$ \\
$\gamma_{AD}$ & 0.179 & 1.01 & eV$\cdot$\AA$^{2}$ \\
\hline
$\gamma_{AC'}$ & 2.443 & 1.08 & eV$\cdot$\AA$^{2}$ \\
$\gamma_{AD'}$ & 0.364 & 0.14 & eV$\cdot$\AA$^{2}$ \\
\hline
$\chi_{AB}$ & 3.688 & 2.41 & eV$\cdot$\AA \\
$\chi_{AC}$ & 2.208 & 2.84 & eV$\cdot$\AA \\
\hline
$\chi_{AC'}$ & 2.071 & 1.09 & eV$\cdot$\AA \\
\hline\hline
\end{tabular}
\end{table}

By comparing both five-hopping\cite{Milton, DuarteLuan} and ten-hopping models, one can notice that the continuum approximated structure factors (Eqs.~(\ref{31})-(\ref{36})) have the same form in both models, and consequently the BP Hamiltonians in both models are also similar within the long-wavelength approach. Thus, all the complicated contributions due to the long-range hoppings are translated to different values of the extended structure coefficients. Therefore, the electronic properties derived by the continuum approximation for both models are qualitatively equivalents. This is a very important issue, because relevant works on the theory of BP systems based on the five-hopping long-wavelength approximation have been already reported\cite{Milton,DuarteLuan}, and here we are showing that their results are still qualitatively valid.

\subsection{Monolayer phosphorene}\label{subsec.ContinuumMono}

The long-wavelength Hamiltonian for the monolayer BP based on the ten-hopping description for low-energy contribution $H_{k}^{+}$ is given by
\begin{equation}
H_{k}^{+} \hspace{-0.075cm}=\hspace{-0.075cm} \left(\hspace{-0.1cm}
\begin{array}{cc}
u_0 \hspace{-0.1cm}+\hspace{-0.1cm} \eta_x k_x^2 \hspace{-0.1cm}+\hspace{-0.1cm} \eta_y k_y^2 & \delta \hspace{-0.1cm}+\hspace{-0.1cm}\gamma_x k_x^2 \hspace{-0.1cm}+\hspace{-0.1cm} \gamma_y k_y^2 \hspace{-0.1cm}+\hspace{-0.1cm} i\chi k_y \\
\delta \hspace{-0.1cm}+\hspace{-0.1cm}\gamma_x k_x^2 \hspace{-0.1cm}+\hspace{-0.1cm} \gamma_y k_y^2 \hspace{-0.1cm}-\hspace{-0.1cm} i\chi k_y & u_0 \hspace{-0.1cm}+\hspace{-0.1cm} \eta_x k_x^2 \hspace{-0.1cm}+\hspace{-0.1cm} \eta_y k_y^2 
\end{array}
\right),
\label{eq37}
\end{equation}
which has exactly the same form as the one corresponding to the five-hopping model\cite{Milton, DuarteLuan}, except for the constants values. For both models, the values of the coefficients in Eq.~(\ref{eq37}) are given by $u_0 = \delta_{AA} + \delta_{AD}$, $\eta_{x} = \eta_{AA} + \eta_{AD}$, $\eta_{y} = \gamma_{AA} + \gamma_{AD}$, $\delta = \delta_{AB} + \delta_{AC}$, $\gamma_x = \eta_{AB} + \eta_{AC}$, $\gamma_y = \gamma_{AB} + \gamma_{AC}$, and $\chi = \chi_{AB} + \chi_{AC}$. By diagonalizing the Hamiltonian (\ref{eq37}), one can obtain the dispersion relations for electrons and holes as
\begin{equation}
E_{k}^{+} \hspace{-0.05cm}=\hspace{-0.05cm} u_0 \hspace{-0.05cm}+\hspace{-0.05cm} \eta_x k_{x}^{2} \hspace{-0.05cm}+\hspace{-0.05cm} \eta_y k_{y}^{2} \hspace{-0.05cm}\pm\hspace{-0.05cm} \sqrt{(\delta \hspace{-0.05cm}+\hspace{-0.05cm} \gamma_x k_{x}^{2} \hspace{-0.05cm}+\hspace{-0.05cm} \gamma_y k_{y}^{2})^{2} \hspace{-0.05cm}+\hspace{-0.05cm} \chi^{2}k_{y}^{2}},
\label{eq38}
\end{equation}
where the plus (minus) sign yields the conduction (valance) band. This leads to an energy gap of $E_g = 2\delta \approx 1.838$ eV that is consistent with Ref.~[\onlinecite{Rudenko1}]. 

\subsection{Bilayer phosphorene}\label{subsec.ContinuumBi}

Analogously to the monolayer case, we can derive a long-wavelength Hamiltonian describing the lowest energy bands close to the Fermi level of bilayer BP as
\begin{align}\label{eq39}
& H_0 + H_2 \pm H_3/2 =\nonumber\\  
&\hspace{-0.075cm} \left(\hspace{-0.1cm}
\begin{array}{cc}
u_0^{\pm} \hspace{-0.1cm}+\hspace{-0.1cm} \eta_x^{\pm} k_x^2 \hspace{-0.1cm}+\hspace{-0.1cm} \eta_y^{\pm} k_y^2 & \delta^{\pm} \hspace{-0.1cm}+\hspace{-0.1cm}\gamma_x^{\pm} k_x^2 \hspace{-0.1cm}+\hspace{-0.1cm} \gamma_y^{\pm} k_y^2 \hspace{-0.1cm}+\hspace{-0.1cm} i\chi^{\pm}k_y \\
\delta^{\pm} \hspace{-0.1cm}+\hspace{-0.1cm}\gamma_x^{\pm} k_x^2 \hspace{-0.1cm}+\hspace{-0.1cm} \gamma_y^{\pm} k_y^2 \hspace{-0.1cm}-\hspace{-0.1cm} i\chi^{\pm} k_y & u_0^{\pm} \hspace{-0.1cm}+\hspace{-0.1cm} \eta_x^{\pm} k_x^2 \hspace{-0.1cm}+\hspace{-0.1cm} \eta_y^{\pm} k_y^2 
\end{array}\right),
\end{align}
where $u_0^{\pm} = u_0 \pm \delta_{AD'}/2$, $\eta_{x}^{\pm} = \eta_{x} \pm \eta_{AD'}/2$, $\eta_{y}^{\pm} = \eta_{y} \pm \gamma_{AD'}/2$, $\delta^{\pm} = \delta \pm \delta_{AC'}/2$, $\gamma_x^{\pm} = \gamma_x \pm \eta_{AC'}/2$, $\gamma_y^{\pm} = \gamma_y \pm \gamma_{AC'}/2$, and $\chi^{\pm} = \chi \pm \chi_{AC'}/2$. One can clearly see that the low-energy Hamiltonian for bilayer BP (\ref{eq39}) has exactly the same structure as the monolayer Hamiltonian (\ref{eq37}) for the case of zero bias ($\Delta = 0$) in the long-wavelength limit, differing only by the coefficient values of each matrix element. The low-energy bands obtained from Eq.~(\ref{eq39}) at the $\Gamma$ point are given by Eqs.~(\ref{eq11}) and (\ref{eq12}) with 
\begin{subequations}
\begin{align}
& \epsilon_{1}^{\pm} = u_0^{\pm} + \eta_{x}^{\pm}k_x^{2} + \eta_{y}^{\pm} k_{y}^{2} \label{eq40a}\\
& \epsilon_{2}^{\pm} = \sqrt{(\delta^{\pm} + \gamma_x^{\pm}k_x^{2} + \gamma_y^{\pm} k_y^{2})^{2} + (\chi^{+} k_y)^{2}} \label{eq40b}
\end{align}
\end{subequations}

For the zero bias case ($\Delta = 0$), the energy levels are given by $\epsilon_1^{\pm} +s \epsilon_2^{\pm}$, i.e.
\begin{equation}
E_{s}^{\pm} \hspace{-0.05cm}=\hspace{-0.05cm} u_0^{\pm} \hspace{-0.05cm}+\hspace{-0.05cm} \eta_x^{\pm} k_{x}^{2} \hspace{-0.05cm}+\hspace{-0.05cm} \eta_y^{\pm} k_{y}^{2} \hspace{-0.05cm}+\hspace{-0.05cm} s \sqrt{(\delta^{\pm} \hspace{-0.05cm}+\hspace{-0.05cm} \gamma_x^{\pm} k_{x}^{2} \hspace{-0.05cm}+\hspace{-0.05cm} \gamma_y^{\pm} k_{y}^{2})^{2} \hspace{-0.05cm}+\hspace{-0.05cm} (\chi^{\pm}k_{y})^{2}},\label{eq41}
\end{equation}
with $s = \pm 1$, where the positive (negative) sign denotes the conduction (valence) bands. Eq.~(\ref{eq41}) has exactly the same structure as the energy bands of the monolayer case Eq.~(\ref{eq38}), as already expected.

\subsection{N-layer phosphorene}\label{subsec.ContinuumNlayer}

As we can anticipate from the previous subsections, the Hamiltonian for the $N$-layer case in the continuum approximation should be composed of $N$ blocks of monolayer type Hamiltonians with the corresponding modified coefficients. Therefore, we can write the low-energy Hamiltonians in the continuum approximation as
\begin{equation}
\left(\hspace{-0.05cm}
\begin{array}{cc}
u_0^{n} \hspace{-0.05cm}+\hspace{-0.05cm} \eta_x^{n} k_x^2 \hspace{-0.05cm}+\hspace{-0.05cm} \eta_y^{n} k_y^2 & \delta^{n} \hspace{-0.05cm}+\hspace{-0.05cm}\gamma_x^{n} k_x^2 \hspace{-0.05cm}+\hspace{-0.05cm} \gamma_y^{n} k_y^2 \hspace{-0.05cm}+\hspace{-0.05cm} i\chi^{n} k_y \\
\delta^{n} \hspace{-0.05cm}+\hspace{-0.05cm}\gamma_x^{n} k_x^2 \hspace{-0.05cm}+\hspace{-0.05cm} \gamma_y^{n} k_y^2 \hspace{-0.05cm}-\hspace{-0.05cm} i\chi^{n} k_y & u_0^{n} \hspace{-0.05cm}+\hspace{-0.05cm} \eta_x^{n} k_x^2 \hspace{-0.05cm}+\hspace{-0.05cm} \eta_y^{n} k_y^2 
\end{array}\hspace{-0.05cm}\right),
\label{eq42}
\end{equation}
with $u_0^{n} = u_0 + \lambda_{n} \delta_{AD'}$, $\eta_{x}^{n} = \eta_{x} + \lambda_{n} \eta_{AD'}$, $\eta_{y}^{n} = \eta_{y} + \lambda_{n} \gamma_{AD'}$, $\delta^{n} = \delta + \lambda_{n} \delta_{AC'}$, $\gamma_x^{n} = \gamma_x + \lambda_{n} \eta_{AC'}$, $\gamma_y^{n} = \gamma_y + \lambda_{n} \gamma_{AC'}$, and $\chi^{n} = \chi + \lambda_{n} \chi_{AC'}$, where $\lambda_n = \cos(n\pi/(N + 1))$. Thus, we have reduced the $N$-layer BP problem to an effective monolayer BP system with layer-dependent coefficients. 

Assuming the limit $N \rightarrow \infty$, i.e. at the bulk BP regime, we can write
\begin{equation}
\phi_i \propto \sin\left(i\frac{n\pi}{N+1}\right) = \sin\left(id\frac{n\pi}{d(N+1)}\right),
\label{eq43}
\end{equation}
where $d\approx 10.69$ \AA\ is the lattice parameter along the $z$-direction for the $AB$-stacked case (see Fig.~\ref{Fig1}(c)). \cite{Peeters, Rahaman} Rewriting Eq.~(\ref{eq43}) as $\phi_i \propto \sin(k_{z}z)$, with $z = jd$ and $k_{z} = n\pi/d(N+1)$, we can obtain the band structure in terms of $k_z$ and consider only terms up to second order within the long-wavelength approximation. It is important to point out that the long-wavelength approximation for the bulk case is only valid for small $k_{z}$, which means $n\pi/(N + 1) \ll 1 $. However, the lowest energy bands occur for $n = N$, such that the inequality can not be satisfied. This issue can be easily figured out by using the cosine identity $\cos(N\pi/(N + 1)) = -\cos(\pi/(N + 1))$, and thus avoiding inconsistencies within the long-wavelength approximation. Taking that in account, we can write the low-energy bands as
\begin{align}
E_{s}^{n} &= \bar{u_0} + \bar{\eta_x} k_{x}^{2} + \bar{\eta_y} k_{y}^{2} + \bar{\eta_z} k_{z}^{2} \nonumber \\
& + s \sqrt{(\bar{\delta} + \bar{\gamma_x} k_{x}^{2} + \bar{\gamma_y} k_{y}^{2} + \bar{\gamma_z} k_{z}^{2})^{2} + (\bar{\chi}k_{y})^{2}},
\label{eq44}
\end{align}
where $\bar{u_0} = u_0 - \delta_{AD'}$, $\bar{\eta_{x}} = \eta_{x} - \eta_{AD'}$, $\bar{\eta_{y}} = \eta_{y} - \gamma_{AD'}$, $\bar{\delta} = \delta - \delta_{AC'}$, $\bar{\gamma_x} = \gamma_x- \eta_{AC'}$, $\bar{\gamma_y} = \gamma_y- \gamma_{AC'}$, $\bar{\chi} = \chi - \chi_{AC'}$, $\bar{\eta_{z}} = \delta_{AD'}d^{2}/2$ and  $\bar{\gamma_{z}} = \delta_{AC'}d^{2}/2$. From the spectrum Eq.~(\ref{eq44}), one can estimate the effective masses of electrons ($s = +1$) and holes ($s = -1$) along the $z$ direction as
\begin{equation}
m_{z}^{e,h} = \frac{\hbar^{2}}{2(\bar{\eta_{z}} \pm \bar{\gamma_z})}.
\label{eq45}
\end{equation}
The resulting effective masses are $m_{z}^{e} \approx 0.115 m_0$ and $m_{z}^{h} \approx 0.158 m_0$, with $m_0$ being the mass of a free electron. Cyclotron resonance experiments\cite{Narita, Narita1, Morita} on bulk BP found on out-of-plane electron and hole effective masses as $m_{z}^{e} \approx 0.128 m_0$ and $m_{z}^{h} \approx 0.280 m_0$, respectively. Therefore, the effective masses found here within our continuum model are consistent with experimental measurements\cite{Narita, Narita1, Morita} and also with theoretically predicted values reported in Refs.~[\onlinecite{Morita, TonyMass}], where the discrepancies are possibly due to the slight differences in the out-of-plane lattice constant. Some theoretical papers\cite{Roberto, Tony} have adopted averages of experimental\cite{Narita} and theoretical\cite{TonyMass} values, assuming $m_{z}^{e} = 0.2m_0$ and $m_{z}^{h} = 0.4m_0$ for the electron and hole out-of-plane masses, respectively. We can observe from Eqs.~(\ref{eq44}) and (\ref{eq45}) that the properties of BP in the $z$-direction are more similar to the properties along the $x$-direction than the $y$-direction, since there is no linear term in $k_{z}$. 

\begin{figure}[t]
\centerline{\includegraphics[width = 0.9\linewidth]{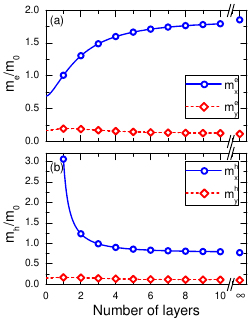}}
\caption{The effective masses in units of free electron mass ($m_0$) along $x$ (solid curves) and $y$ (dashed curves) direction for (a) electrons and (b) holes as a function of the number of layers. The open symbols highlights the behavior of the analytical expression (\ref{eq46}) for integer values of the number of layers $N$.} 
\label{Fig4}
\end{figure}

We can also investigate how the effective masses along the $x$ and $y$ directions change with the number of layers. According to Ref.~[\onlinecite{Milton}], one can estimate the effective masses for the $n$-th sub-Hamiltonian in the $N$-layer BP in a similar way as in monolayer case\cite{Milton}, that reads
\begin{equation}
m_{x}^{e,h} \hspace{-0.05cm}=\hspace{-0.05cm} \frac{\hbar^{2}}{2(\eta_{x}^{n} \hspace{-0.05cm}\pm\hspace{-0.05cm} \gamma_{x}^{n})}, \quad m_{y}^{e,h} \hspace{-0.05cm}=\hspace{-0.05cm} \frac{\hbar^{2}}{2(\eta_{x}^{n} \hspace{-0.05cm}\pm\hspace{-0.05cm} \gamma_{x}^{n} \hspace{-0.05cm}\pm\hspace{-0.05cm} (\chi^{n})^{2}/2\delta^{n})},
\label{eq46}
\end{equation}
where the coefficients are layer-dependent. Fig.~\ref{Fig4} is a plot of the effective masses in units of $m_0$ along $x$ and $y$ directions for (a) electrons and (h) holes as a function of the number of layers. One can notice the effective masses $m_x^{e,h}$ along the $x$-direction are more sensitive to changes with respect to the number of layer than $m_y^{e,h}$. At the limit $N \rightarrow \infty$, the effectives masses for electrons and holes in both in-plane directions converge to values of bulk BP: $m_{x}^{e} \approx 1.855m_0$, $m_{x}^{h} \approx 0.774m_0$, $m_{y}^{e} \approx 0.115m_0$ and $m_{y}^{h} \approx 0.104m_0$, respectively. Recent works\cite{Tony, TonyMass} in multilayer BP within the ${\mathbf k} \cdot {\mathbf p}$ model have assumed the following effective masses $m_{x}^{e} \approx 0.7m_0$, $m_{x}^{h} \approx 1.0 m_0$, and $m_{y}^{e} = m_{y}^{h} \approx 0.08m_0$ by taking average values between experimental and theoretical results,\cite{Narita, Narita1, Morita} but in fact these values are close to the ones for monolayer BP.\cite{Milton} 

\section{Landau Levels}\label{sec.LLs}

\begin{figure}[b]
\centerline{\includegraphics[width = \linewidth]{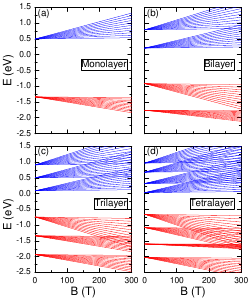}}
\caption{(Color online) Landau levels for electrons (blue curves) and holes (red curves) as a function of perpendicular magnetic field for (a) monolayer, (b) bilayer, (c) trilayer, and (d) tetralayer BP. It is shown the first eleven ($n = 0$, $1$, $\cdots$, $10$) lowest states for each subband.} 
\label{Fig5}
\end{figure} 

In order to exemplify the physics of the analytical model developed here, we discuss the influence of an external and uniform magnetic field perpendicular to the BP sheets (${\mathbf B} = B\hat{z}$), obtaining the Landau levels for $N$-layer BP. By considering the Peierls substitution ${\mathbf p}\rightarrow{\mathbf p}-e{\mathbf A}$ into the continuum Hamiltonian (\ref{eq42}) and using the Landau gauge ${\mathbf A} = (-By,0,0)$, we can readily generalize the Landau level expression for the multilayer BP case by following the same straightforward procedure adopted in Ref.~[\onlinecite{Milton}], such as 
\begin{equation}\label{eqlls}
E_{e,h}^{n} = u_0^{n} \pm \delta^{n} \pm \hbar\omega_{e,h}\left(n + 1/2\right),
\end{equation}
with frequency defined as
\begin{equation}\label{eqW}
\omega_{e,h} = \frac{eB}{ \sqrt{ m^{e,h}_{x} m^{e,h}_{y} } },
\end{equation}
where the general layer-dependent effective masses are given by Eq.~(\ref{eq46}) and the sign $+$ ($-$) corresponds to the electron $e$ (hole $h$) branches. Note that the spectrum obtained from Eq.~(\ref{eqlls}) has a linear dependence on $B$, similarly to conventional $2$D electron gas spectrum, i.e. the dispersion is typical of Schr\"odinger Fermions. In Fig.~\ref{Fig5}, we show the Landau level spectra for electrons (blue lines) and holes (red lines) as a function of perpendicular magnetic field for (a) one, (b) two, (c) three and (d) four BP layers. One can notice that all subbands for both electron and hole branches increase linearly but with different slopes due to different anisotropic effective masses and cross at some high magnetic amplitude. Similar results for multilayer BP within the ${\mathbf k} \cdot {\mathbf p}$ method was already reported in Ref.~[\onlinecite{Tony}].

\begin{figure}[b]
\centerline{\includegraphics[width = \linewidth]{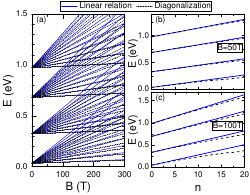}}
\caption{(Color online) Landau levels for electrons of tetralayer BP ($N=4$) as a function of (b) perpendicular magnetic field and (b, c) Landau energy index $n$ with different magnetic field. A comparison between the energy bands obtained by using the linear dispersion Eq.~(\ref{eqlls}) (blue solid curves) and the diagonalization of Hamiltonian (\ref{eq42}) with a perpendicular magnetic field (black dashed curves) is shown. It was assumed just the first eleven low Landau levels for each electronic subband in panel (a).} 
\label{Fig5b}
\end{figure}

Aiming to check the validity of the analytical results obtained via Eq.~(\ref{eqlls}), we plot in Fig.~\ref{Fig5b} the electronic Landau energy branches for tetralayer BP obtained by using the linear dispersion Eq.~(\ref{eqlls}) (blue solid curves) and the diagonalization of Hamiltonian (\ref{eq42}) with a perpendicular magnetic field solved numerically (black dashed curves). Panel (a) shows a plot of the Landau levels as function of $B$ and in panels (b) and (c) we depict the results for the Landau levels versus the Landau energy index $n$ for $B=50$ T and $B=100$ T, respectively. As shown in Fig.~\ref{Fig5b}(a), the approximate linear relation describes very appropriately the Landau energies even at high magnetic field values. It can be observed a small deviation from the linear dependence on large field and high energy index $n$ for each subband. This is due to interband coupling terms depending on $\chi$ that correspond to the off-diagonal matrix elements of Eq.~(\ref{eq42}). Figs.~\ref{Fig5b}(b) and \ref{Fig5b}(c) confirms the good accordance between these results even for high Landau energy index $n$. One can also observe clearly from panels \ref{Fig5b}(b) and \ref{Fig5b}(c) that the Landau levels are equally spaced and have also a linear dependence on $n$. Therefore, the excellent agreement between those results demonstrates that the approximate linear relation for the Landau levels (\ref{eqlls}) are able to describe accurately the main features even at high magnetic regime and low-energy indexes. The valid of the linear approximation for the electron and holes energy branches was also discussed in Refs.~[\onlinecite{Zhou, Tony, Milton}]. According to them, the Landau levels obey with a good agreement the linear dependence at low magnetic field values (up to $\lessapprox50$ T).

\section{CONCLUSIONS}\label{sec.Conclusions}

In summary, we have studied the electronic properties of multilayer BP and analytically derived an effective model for this system with arbitrary number of BP layers, based on a recently proposed tight-binding approach utilizing ten intralayer and four interlayer hopping parameters. We have shown that a decomposition into $N$ effective Hamiltonians of order $2$ naturally emerges from the $N$ coupled BP problem for the low energy bands, by mapping the complicated problem of $N\times N$ tridiagonal hermitian $4\times4$ blocks to a system of $N$ uncoupled single layer BP. By using the advantage of the sublattice symmetries between $A/D$ and $B/C$ atomic sites as a consequence of the $D_{2h}$ group invariance of the BP lattice, we have separated the Hamiltonians that describe low and high energy bands, associating them with bond and anti-bond wavefunction amplitudes, respectively. We have verified that the low and high energy bands are described by the sum $\psi_{AB,i} + \psi_{DC,i}$ and difference $\psi_{AB,i} - \psi_{DC,i}$ of the probability amplitudes of the equivalent sublattices in each layer $i$. Using the decoupled tight-binding model for multilayer BP, we have expanded the structure factor around the $\Gamma$ point up to second order of the matrix elements of BP Hamiltonian in order to achieve a long-wavelength approximation for the system. This has allowed us to obtain the dispersion relations for electrons and holes in the vicinity of the Fermi level, as well as general expressions for gap energy and effective masses with an explicit dependence on the number of BP layers. Our findings have shown that the effective continuum model displays good agreement with previous results of both first-principles calculations and tight-binding approximation reported in the literature by Rudenko et al.\cite{Rudenko, Rudenko1}, reproducing well the band structures near the Fermi level. Moreover, in the limit of large number of BP layers ($N\rightarrow\infty$, i.e. bulk BP), we found an energy band gap of $\approx0.414$ eV with our simple multilayer continuum model. This is consistent with the values obtained by previous first-principles calculations\cite{DFTgap} ($\approx 0.43$ eV) and tight-binding model\cite{Rudenko, Rudenko1} ($\approx 0.40$ eV). As an example of the application of the model, we have considered the case of a perpendicular magnetic field to multilayer BP and found general expression for the electron and holes Landau level spectra. Therefore, the analytical analysis developed in the present paper captures the essential physics of multilayer BP and is suitable for large-scale investigation, since it obtains accurate quantitative results and is less computationally demanding than numerical tight-binding model and first-principles calculations for large BP systems.

\section*{ACKNOWLEDGMENTS} 

This work was financially supported by the Brazilian Council for Research (CNPq), under the PRONEX/FUNCAP and CAPES foundation.

\appendix
\section*{APPENDIX}
\subsection{Structure factors}\label{appA}

\begin{table}[b]
\label{table1}
\caption{Intralayer ($t_i$) and interlayer ($t_i^{\perp}$) hopping parameters from Ref.~[\onlinecite{Rudenko1}].}
\centering 
\begin{tabular}{cccc}
\hline\hline
Parameter \hspace{0.1cm}& \hspace{0.1cm} Value (eV)\hspace{0.1cm} & \hspace{0.1cm} Parameter \hspace{0.1cm} & \hspace{0.1cm}Value (eV) \\
\hline
$t_{1}$ & $-1.486$ & $t_{2}$ & $3.729$ \\
$t_{3}$ & $-0.252$ & $t_{4}$ & $-0.071$ \\
$t_{5}$ & $-0.019$ & $t_{6}$ & $0.186$  \\
$t_{7}$ & $-0.063$ & $t_{8}$ & $0.101$  \\
$t_{9}$ & $-0.042$ & $t_{10}$ & $0.073$ \\
\hline
$t_1^{\perp}$ & $0.524$ & $t_2^{\perp}$ & $0.180$ \\
$t_3^{\perp}$ & $-0.123$ & $t_4^{\perp}$ & $-0.168$ \\
\hline\hline
\end{tabular}
\end{table}

\begin{figure*}[t]
\centerline{\includegraphics[width = \linewidth]{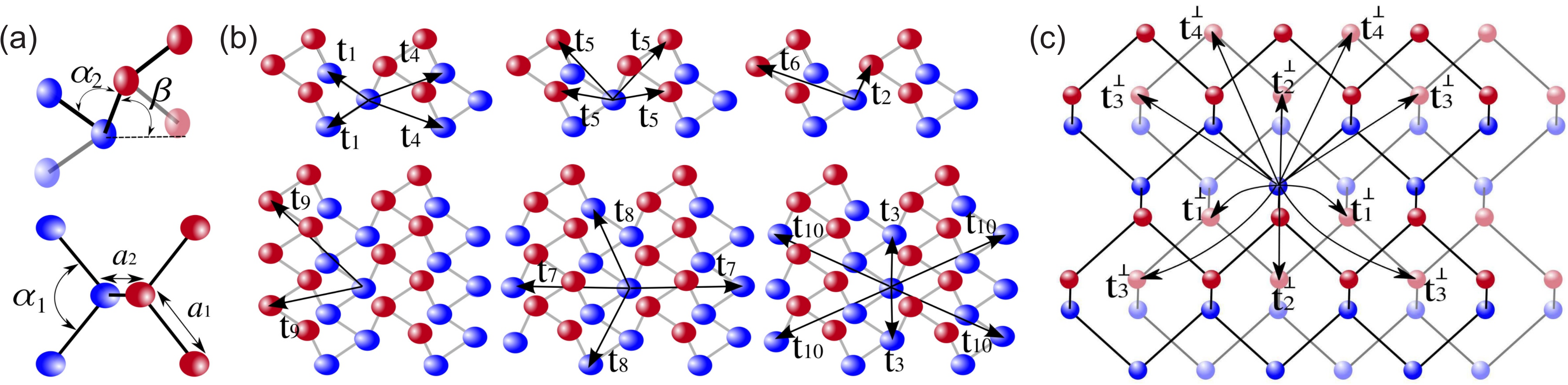}}
\caption{(Color online) Lattice structure of multilayer BP systems and parameters for the tight-binding model. (a) Definitions of the bond lengths and bond angles. Sketches of the (b) ten intralayer and (c) four interlayer hopping parameters.} 
\label{FigAppendix}
\end{figure*}

Here, we shall derive the structure factors corresponding to the matrix elements of the ten-hopping tight-binding Hamiltonian considered in our approach. Figure~\ref{FigAppendix}(a) shows the lattice structure of multilayer BP systems, emphasizing the the bond lengths and bond angles between the phosphorus atomic sites, where $a_1=2.22$\AA\ ($a_2=2.24$\AA) is the distance between nearest-neighbor sites in sublattices $A$ and $B$ or $C$ and $D$ ($A$ and $C$ or $B$ and $D$), $\alpha_1 = 96.5^{\circ}$, $\alpha_2 = 101.9^{\circ}$,  and $\beta=72^{\circ}$. Figs.~\ref{FigAppendix}(b) and \ref{FigAppendix}(c) indicate the ten-intralayer $t_i$ and four-interlayer $t_i^{\perp}$ hopping parameters for the tight-binding model, respectively. The hopping energies are depicted in Table II. We shall just show how to compute the expression for $t_{AB}$, since the other terms $t_{AA}$, $t_{AC}$, $t_{AD}$, $t_{AC'}$ and $t_{AD'}$ can be obtained in an analogous way. The $t_{AB}$ term corresponds to all the terms involving the coupling energies between $A-B$ and $C-D$. Therefore, by analyzing Fig.~\ref{FigAppendix}(b), we have
\begin{align}
\hspace{-0.075cm}\mathcal{H}_{AB} \hspace{-0.05cm}&=\hspace{-0.05cm} \sum_{i,j}t_{ij}(a_i^\dag b_j \hspace{-0.05cm}+\hspace{-0.05cm} d_i^\dag c_j) \hspace{-0.05cm}+\hspace{-0.05cm} h.c.\hspace{-0.05cm}=\hspace{-0.05cm} t_1\sum_{i,j}(a_i^\dag b_j \hspace{-0.05cm}+\hspace{-0.05cm} d_i^\dag c_j)\nonumber\\
&+\hspace{-0.05cm} t_4\sum_{i,j}(a_i^\dag b_j \hspace{-0.05cm}+\hspace{-0.05cm} d_i^\dag c_j) \hspace{-0.05cm}+\hspace{-0.05cm} t_8\sum_{i,j}(a_i^\dag b_j \hspace{-0.05cm}+\hspace{-0.05cm} d_i^\dag c_j) \hspace{-0.05cm}+\hspace{-0.05cm} h.c.
\label{eq47}
\end{align}
Labeling $t_1$, $t_4$ and $t_8$ by $t_n$, then each part of Eq.~(\ref{eq47}) can be calculated as
\begin{align}
& t_n\sum_{i,j}(a_i^\dag b_j \hspace{-0.05cm}+\hspace{-0.05cm} d_i^\dag c_j) \hspace{-0.05cm}=\hspace{-0.05cm} \frac{t_n}{N}\hspace{-0.075cm}\sum_{i,j,k,k'}\hspace{-0.075cm}(a_k^\dag b_{k'} \hspace{-0.05cm}+\hspace{-0.05cm} d_k^\dag c_{k'})e^{-i\vec{k}\cdot\vec{r_i}}e^{i\vec{k'}\cdot\vec{r_j}} \nonumber\\
&\hspace{1.7cm}=\hspace{-0.05cm} \frac{t_n}{N}\hspace{-0.075cm}\sum_{i,j,k,k'}\hspace{-0.075cm}(a_k^\dag b_{k'} \hspace{-0.05cm}+\hspace{-0.05cm} d_k^\dag c_{k'})e^{i\vec{k}\cdot(\vec{r_i}\hspace{-0.05cm}-\hspace{-0.05cm}\vec{r_j})}e^{i(\vec{k'}-\vec{k})\cdot\vec{r_j}}\nonumber\\
&\hspace{1.7cm}=\hspace{-0.075cm} \sum_k\hspace{-0.075cm} \left(\hspace{-0.05cm}t_n\sum_{m_n} e^{i\vec{k}\cdot\vec{\delta}_m}\hspace{-0.05cm}\right)(a_k^\dag b_{k} \hspace{-0.05cm}+\hspace{-0.05cm} d_k^\dag c_{k}),
\label{eq48}
\end{align}
where $\sum_{m_n}$ denotes a summation over all $\vec{r}_j$ neighbors of $\vec{r}_i$ linked by the hopping $t_n$ and $\vec{\delta}_i = \vec{r}_i-\vec{r}_j$ corresponds to the distance vector between the atomic site $i$ and $j$. Therefore, we arrive at the following expression
\begin{align}
\hspace{-0.125cm}\mathcal{H}_{AB} \hspace{-0.075cm}&=\hspace{-0.075cm} \sum_k \hspace{-0.1cm}\left(\hspace{-0.1cm} t_1\sum_{m_1} e^{i\vec{k}\cdot\vec{\delta}_{m_1}} \hspace{-0.075cm}+\hspace{-0.075cm} t_4\sum_{m_4} e^{i\vec{k}\cdot\vec{\delta}_{m_4}}\hspace{-0.075cm}+\hspace{-0.075cm} t_8\hspace{-0.1cm}\sum_{m_8}\hspace{-0.1cm} e^{i\vec{k}\cdot\vec{\delta}_{m_8}} \hspace{-0.075cm}\right)\nonumber\\
&\hspace{-0.75cm}\times\hspace{-0.075cm}(a_k^\dag b_{k} \hspace{-0.075cm}+\hspace{-0.075cm} d_k^\dag c_{k}) \hspace{-0.075cm}+\hspace{-0.075cm} h.c. \hspace{-0.075cm}=\hspace{-0.1cm} \sum_k \hspace{-0.1cm}t_{AB}(k)(a_k^\dag b_{k} \hspace{-0.075cm}+\hspace{-0.075cm} d_k^\dag c_{k}) \hspace{-0.075cm}+\hspace{-0.075cm} h.c.
\label{eq49}
\end{align}

By replacing the distance vectors, calculated according to Fig.~\ref{FigAppendix}, into Eq.~(\ref{eq49}), one can find
\begin{align}
\hspace{-0.25cm}t_{AB}(k) \hspace{-0.05cm}&=\hspace{-0.05cm} 2t_{1}\cos\left[a_1 \sin(\alpha_1 /2)k_x\right]e^{-ia_1 \cos(\alpha_1 / 2)k_y}\nonumber\\
&+\hspace{-0.05cm} 2t_{4}\cos\left[a_1\sin(\alpha_1 /2)k_x\right]e^{i[2a_2 \cos\beta + a_1 \cos(\alpha_1 / 2)]k_y} \nonumber\\
&+\hspace{-0.05cm}2t_{8}\cos\left[3a_1 \sin(\alpha_1 /2)k_x\right]e^{-ia_1 \cos(\alpha_1 / 2)k_y}.
\label{eq50}
\end{align}
Analogously, we can obtain the intralayer ($t_{AA}$, $t_{AC}$ and $t_{AD}$) and interlayer ($t_{AD'}$ and $t_{AC'}$) coupling contributions, such as
\begin{subequations}
\begin{align}
\hspace{-0.25cm}t_{AA}(k) \hspace{-0.05cm}&=\hspace{-0.05cm}  2t_{3}\cos\left[2a_1 \sin(\alpha_1 /2)k_x\right] \nonumber\\
& + \hspace{-0.05cm}2t_{7}\cos\{\left[2a_1 \cos(\alpha_1 /2) + 2a_2\cos\beta\right]k_y\}\nonumber\\
& + \hspace{-0.05cm}4t_{10}\cos\left[2a_1 \sin(\alpha_1 /2)k_x\right]\nonumber \\ 
& \times \hspace{-0.05cm}\cos\{\left[2a_1 \cos(\alpha_1 /2) + 2a_2\cos\beta\right]k_y\},\label{eq51a}\\
\hspace{-0.25cm}t_{AC}(k) \hspace{-0.05cm}&=\hspace{-0.05cm} t_{2}e^{ia_2 \cos(\beta)k_y} + t_{6}e^{-i[a_2 \cos\beta + 2a_1 \cos(\alpha_1 / 2)]k_y}\nonumber \\
& + \hspace{-0.05cm}2t_{9}\hspace{-0.025cm}\cos\hspace{-0.025cm}\left[\hspace{-0.025cm}2a_1 \sin\hspace{-0.025cm}(\hspace{-0.025cm}\alpha_1 /2\hspace{-0.025cm})\hspace{-0.025cm}k_x\hspace{-0.025cm}\right]\hspace{-0.05cm}e^{-i\hspace{-0.025cm}[\hspace{-0.025cm}a_2 \cos\beta + 2a_1 \cos\hspace{-0.025cm}(\hspace{-0.025cm}\alpha_1 / 2\hspace{-0.025cm})\hspace{-0.025cm}]\hspace{-0.025cm}k_y},\label{eq51a}\\
\hspace{-0.25cm}t_{AD}(k) \hspace{-0.05cm}&=\hspace{-0.05cm} 4t_{5}\cos\hspace{-0.025cm}\left[\hspace{-0.025cm}a_1 \sin\hspace{-0.025cm}(\hspace{-0.025cm}\alpha_1 /2\hspace{-0.025cm})k_x\hspace{-0.025cm}\right]\hspace{-0.025cm}\nonumber \\ 
& \times \hspace{-0.05cm}\cos\hspace{-0.025cm}\{\hspace{-0.025cm}\left[\hspace{-0.025cm}a_1 \cos\hspace{-0.025cm}(\hspace{-0.025cm}\alpha_1 /2\hspace{-0.025cm}) \hspace{-0.05cm}+\hspace{-0.05cm} a_2\cos\beta\hspace{-0.025cm}\right]\hspace{-0.025cm}k_y\hspace{-0.025cm}\},\label{eq51c}\\
\hspace{-0.25cm}t_{AD'}(k) \hspace{-0.05cm}&=\hspace{-0.05cm} \{ 4t_{3}^{\perp} \cos[2a_1\sin(\alpha_1/2)k_x] \hspace{-0.05cm}+\hspace{-0.05cm} 2t_{2}^{\perp} \} \nonumber \\
& \times \hspace{-0.05cm} \cos\{[a_1\sin(\alpha_1/2) \hspace{-0.05cm}+\hspace{-0.05cm} a_2 \cos(\beta)]k_y\},\label{eq52a}\\
\hspace{-0.25cm}t_{AC'}(k) \hspace{-0.05cm}&=\hspace{-0.05cm} \{ 2t_{1}^{\perp}e^{i a_2 \cos(\beta)k_y} \hspace{-0.025cm}+\hspace{-0.025cm} 2t_4^{\perp}e^{-i [2a_1\sin(\alpha_1/2) \hspace{-0.025cm}+\hspace{-0.025cm} a_2 \cos(\beta)]k_y}\}\nonumber \\
& \times \hspace{-0.05cm}\cos[2a_1\sin(\alpha_1/2)k_x]. \label{eq52b}
\end{align}
\end{subequations}

\subsection{Unitary transformation}\label{appB}
In order to obtain the Hamiltonian (\ref{eq9}), one can apply the following unitary transformation to the bilayer BP Hamiltonian (\ref{eq7}):
\begin{equation}
U = \frac{1}{2}\left(
\begin{array}{cccc}
\mathds{1} & \mathds{1} & \mathds{1} & \mathds{1} \\
\mathds{1} & \mathds{1} & -\mathds{1} & -\mathds{1} \\
-i\mathds{1} & i\mathds{1} & -i\mathds{1} & i\mathds{1} \\
-i\mathds{1} & i\mathds{1} & i\mathds{1} & -i\mathds{1}
\end{array}\right),
\label{eq53}
\end{equation}
where $\mathds{1}$ denotes the $2\times 2$ unit matrix. Therefore, one obtains
\begin{widetext}
\begin{equation}
UH_{bi} U^\dag = \left(
\begin{array}{cccc}
H_0 + H_2 + H_3 /2 & \Delta/2 & 0 & iH_3/2 \\
\Delta & H_0 + H_2 - H_3/2 & -iH_3/2 & 0 \\
0 & iH_3/2 & H_0 - H_2 - H_3/2 & \Delta/2 \\
-iH_3/2 & 0 & \Delta/2 & H_0 - H_2 + H_3/2
\end{array}\right),
\end{equation}
\label{eq53}
\end{widetext}
In Sec.~\ref{subsec.TBbi}, we use the above transformation to put away the terms appearing in the off-diagonal block.


\begin{references}

\bibitem{CastroNetoReview} 
A. H. Castro Neto, F. Guinea, N. M. R. Peres, K. S. Novoselov, and A. K. Geim, Rev. Mod. Phys. \textbf{81}, 109 (2009).

\bibitem{Misha1}
M. I. Katsnelson, \textit{Graphene: Carbon in Two Dimensions} (Cambridge University Press, 2012).

\bibitem{Geim}
A. K. Geim and I. V. Grigorieva, Nature \textbf{499}, 419 (2013).

\bibitem{silicene}
P. De Padova, C. Ottaviani, C. Quaresima, B. Olivieri, P. Imperatori, E. Salomon, T. Angot, L. Quagliano, C. Romano, A. Vona, M. Muniz-Miranda, A. Generosi, B. Paci, and G. L. Lay, 2D Materials \textbf{1}, 021003 (2014).

\bibitem{germanene}
M. E. D\'avila, L. Xian, S. Cahangirov, A. Rubio, and G. L. Lay, New Jour. of Phys. \textbf{16}, 095002 (2014).

\bibitem{MoS2}
B. Radisavljevic, A. Radenovic, J. Brivio, V. Giacometti, and A. Kis, Nat. Nanotechnol. \textbf{6}, 147 (2011).

\bibitem{TonyBook}
P. Avouris, T. F. Heinz, and T. Low, \textit{2D Materials: Properties and Devices} (Cambridge University Press, 2017).

\bibitem{Two-dimensionalAtomicCrystals}
K. S. Novoselov, D. Jiang, F. Schedin, T. J. Booth, V. V. Khotkevich, S. V. Morozov, and A. K. Geim, Proc. Natl. Acad. Sci. U.S.A. \textbf{102}(30), 10451 (2005). 

\bibitem{bp1}
L. Li, Y. Yu, G. J. Ye, Q. Ge, X. Ou, H. Wu, D. Feng, X. H. Chen, and Y. Zhang, Nat. Nanotech. \textbf{9}, 372 (2014). 

\bibitem{bp2}
H. Liu, A. T. Neal, Z. Zhu, Z. Luo, X. Xu, D. Tom\'anek, and P. D. Ye, ACS Nano \textbf{8}, 4033 (2014).

\bibitem{bp3}
F. Xia, H. Wang, and Y. Jia, Nat. Commun. \textbf{5}, 4458 (2014).

\bibitem{bp4}
S. P. Koenig, R. A. Doganov, H. Schmidt, A. H. Castro Neto, and B. \"Ozyilmaz, Appl. Phys. Lett. \textbf{104}, 103106 (2014).

\bibitem{bp5}
A. Castellanos-Gomez, L. Vicarelli, E. Prada, J. O. Island, K. L. Narasimha-Acharya, S. I. Blanter, D. J. Groenendijk, M. Buscema, G. A. Steele, J. V. Alvarez, H. W. Zandbergen, J. J. Palacios, and H. S. J. van der Zant, 2D Materials \textbf{1}, 025001 (2014). 

\bibitem{bp6}
A. S. Rodin, A. Carvalho, and A. H. Castro Neto, Phys. Rev. Lett. \textbf{112}, 176801 (2014).

\bibitem{bp7}
T. Low, R. Rold\'an, H. Wang, F. Xia, P. Avouris, L. M. Moreno, and F. Guinea, Phys. Rev. Lett. \textbf{113}, 106802 (2014). 

\bibitem{Yuan}
H. Yuan, X. Liu, F. Afshinmanesh, W. Li , G. Xu, J. Sun, B. Lian, A. G. Curto, G. Ye, Y. Hikita, Z. Shen, S.-C Zhang, X. Chen, M. Brongersma, H. Y. Hwang, and Y. Cui, Nat. Nanotechnol. \textbf{10}, 707 (2015).

\bibitem{Fazzio}
Q. Liu, X. Zhang, L. B. Abdalla, A. Fazzio, and A. Zunger, Nano Lett. \textbf{15}, 1222 (2015).

\bibitem{Tran}
V. Tran, R. Soklaski, Y. Liang, and L. Yang, Phys. Rev. B \textbf{89}, 235319 (2014).

\bibitem{Gomez}
A. Castellanos-Gomez, J. Phys. Chem. Lett. \textbf{6}(21), 4280 (2015).

\bibitem{Dolui}
K. Dolui and S. Y. Quek, Sci. Rep. \textbf{5}, 11699 (2015).

\bibitem{Das}
S. Das, W. Zhang,. M.. Demarteau, A. Hoffmann, M. Dubey, and A. Roelofs, Nano Lett. \textbf{14}(10), 5733 (2014).

\bibitem{Kim} 
J. Kim, S. S. Baik, S. H. Ryu, Y. Sohn, S. Park, B.-G. Park, J. Denlinger, Y. Yi, H. J. Choi, and K. S. Kim, Science \textbf{349}, 723 (2015).

\bibitem{Katsnelson} 
S. Yuan, E. van Veen, M. I. Katsnelson, and R. Rold\'an, Phys. Rev. B \textbf{93}, 245433 (2016).

\bibitem{Andrey}
G. Zhang, S. Huang, A. Chaves, C. Song, V. O. \"Oz\c{c}elik, T. Low, and H. Yan, Nat. Commun. \textbf{8}, 14071 (2017).

\bibitem{Rudenko}
A. N. Rudenko and M. I. Katsnelson, Phys. Rev. B \textbf{89}, 201408(R) (2014).

\bibitem{Tran1}
V. Tran and L. Yang, Phys. Rev. B \textbf{89}, 245407 (2014).

\bibitem{Carvalho} 
A. Carvalho, A. S. Rodin, and A. H. Castro Neto, Europhys. Lett. \textbf{108}, 47005 (2014).

\bibitem{DFTgap}
L. Liang, J. Wang, W. Lin, B. G. Sumpter, V. Meunier, and M. Pan, Nano Lett. \textbf{14}, 6400 (2014).

\bibitem{Zhou}
X. Y. Zhou, R. Zhang, J. P. Sun, Y. L. Zou, D. Zhang, W. K. Lou, F. Cheng, G. H. Zhou, F. Zhai, and K. Chang, Sci. Rep. \textbf{5}, 12295 (2015).

\bibitem{Li}
P. Li and I. Appelbaum, Phys. Rev. B \textbf{90}, 115439 (2014).

\bibitem{Roberto}
C. Lin, R. Grassi, T. Low, and A. S. Helmy, Nano Lett. \textbf{16}, 1683 (2016).

\bibitem{Tony} 
Y. Jiang, R. Rold\'an, F. Guinea, and T. Low, Phys. Rev. B \textbf{92}, 085408 (2015).

\bibitem{Rudenko1}
A. N. Rudenko, S. Yuan, and M. I. Katsnelson, Phys. Rev. B \textbf{92}, 085419 (2015).

\bibitem{Milton}
J. M. Pereira and M. I. Katsnelson, Phys. Rev. B \textbf{92}, 075437 (2015).

\bibitem{DuarteLuan} D. J. P. de Sousa, L. V. de Castro, D. R. da Costa, and J. M. Pereira, Phys. Rev. B \textbf{94}, 235415 (2016).

\bibitem{Peeters}
D. \c{C}akir, C. Sevik, and F. M. Peeters, Phys. Rev. B \textbf{92}, 165406 (2015). 

\bibitem{Rahaman}
A. Mukhopadhyay, L. Banerjee, A. Sengupta, and H. Rahaman, J. Appl. Phys. \textbf{118}, 224501 (2015).

\bibitem{Ezawa}
M. Ezawa, New J. Phys. \textbf{16}, 115004 (2014).

\bibitem{PNAS}
X. Ling, H. Wang, S. Huang, F. Xia, and M. S. Dresselhaus, Proc. Natl. Acad. Sci. U.S.A. \textbf{112}(15), 4523 (2015).

\bibitem{DFTgap1}
V. Wang, Y. C. Liu, Y. Kawazoe, and W. T. Geng, J. Phys. Chem. Lett. \textbf{6}, 4876 (2015).

\bibitem{Narita} %exp bp paper
S.-i. Narita, S.-i. Terada, S. Mori, K. Muro, Y. Akahama, and S. Endo, J. Phys. Soc. Jpn. \textbf{52}, 3544 (1983).

\bibitem{Narita1} %exp bp paper
Y. Akahama, S. Endo, and S.-i. Narita, J. Phys. Soc. Jpn. \textbf{52}, 2148 (1983).

\bibitem{Morita} %review-like exp bp paper
A. Morita, Appl. Phys. A \textbf{39}, 227 (1986).

\bibitem{TonyMass}
T. Low, A. S. Rodin, A. Carvalho, Y. Jiang, H. Wang, F. Xia, and A. H. Castro Neto, Phys. Rev. B \textbf{90}, 075434 (2014).

\end{references}
\end{document}